\documentstyle[epsf]{article}          %% 
\catcode`@=11

\@addtoreset{equation}{section}
\catcode`@=12

\begin{document}

\title{{\small\centerline{May 1999 \hfill SINP/TNP/99-19}}
\medskip 
{\bf Particle dark matter~: an overview\thanks{Invited talk at the
conference ``Cosmology: Theory confronts observations'', held at IIT
Kharagpur, January 1999.}}} 

\author{\bf Palash B. Pal\\ 
\normalsize Saha Institute of Nuclear Physics, 1/AF Bidhan-Nagar,
Calcutta 700064, India} 

\date{}
\maketitle 

\begin{abstract} \normalsize\noindent
I discuss some compelling suggestions about particles which could be
the dark matter in the universe, with special attention to
experimental searches for them.
\end{abstract}
\bigskip\bigskip

This is the write-up of a talk given at a conference where I was
probably the only particle physicist. My task was to give an overview
of one area of cosmology which has a symbiotic relation with particle
physics --- viz., the candidates for dark matter.

I could proceed in two ways. One is to give very brief outlines of all
different suggestions (or all I know about) for dark matter candidate
particles. The second is to focus on a few compelling ones, and give
some details of them. After some thought, I chose the second path. 
I will divide this talk into three sections, by three different kinds
of particles which are potential candidates for dark matter.

%%%%%%%%%%%%%%%%%%%%%%%%%
\section{Light neutrinos}
%%%%%%%%%%%%%%%%%%%%%%%%%
Neutrinos interact very feebly, only through weak
interactions. Despite this fact, they were in thermal equilibrium in
the very early universe because the density of particles was much
higher at that time, and so the mean free path (or the average
reaction time) was small. At that stage, their number density
differed from that of the photons only because one is a fermion and
the other is a boson, and the relation was
\begin{eqnarray}
n_\nu = \frac34 n_\gamma \,.
\label{3/4}
\end{eqnarray}
This age of thermal equilibrium ended once the density of the
particles became small enough in an expanding universe, and it
happened when the temperature was about $1$\,MeV. After that, the
neutrinos followed only the overall expansion. The photon number
density also fell by the cube of the scale factor, so the ratio
remained the same. The only exception to this statement occurred when
the temperature dropped somewhat below the electron mass, and $e^+e^-$
annihilations produced new photons. This event increased the photon
number density by a factor of $11/4$, so that in the present universe
\begin{eqnarray}
\Big( n_\nu \Big)_0 = \frac34 \frac4{11} \Big( n_\gamma \Big)_0
\approx 110 \; {\rm cm}^{-3} \,. 
\end{eqnarray}
If the neutrinos are massive, the energy density is obtained by
multiplying this by the mass. The total energy density of the universe
is parametrized by $10^4h^2\Omega\;{\rm eV/cm}^3$, where $h$ is the
Hubble parameter in units of $100\;{\rm
km\,s^{-1}\,{Mpc}^{-1}}$. Thus, the fraction of energy density
contributed by light neutrinos would be
\begin{eqnarray}
F_\nu = {m_\nu n_\nu \over 10^4h^2\Omega\;{\rm eV/cm}^3} = \left(
{m_\nu \over 92\, {\rm eV}} \right) \cdot {1\over h^2\Omega} \,.
\end{eqnarray}
Atmospheric neutrino data indicate neutrino masses of order
$10^{-1}$\,eV. Solar neutrino data indicate even smaller
masses. Although these do not apply for all neutrino species, it
surely is suggestive of the fact that $F_\nu\ll1$.

%%%%%%%%%%%%%%%%%%%%%%%%%
\begin{figure}
\begin{center}
\centerline{\epsfysize=.3\textheight %\epsfxsize=.6\textwidth
\epsfbox{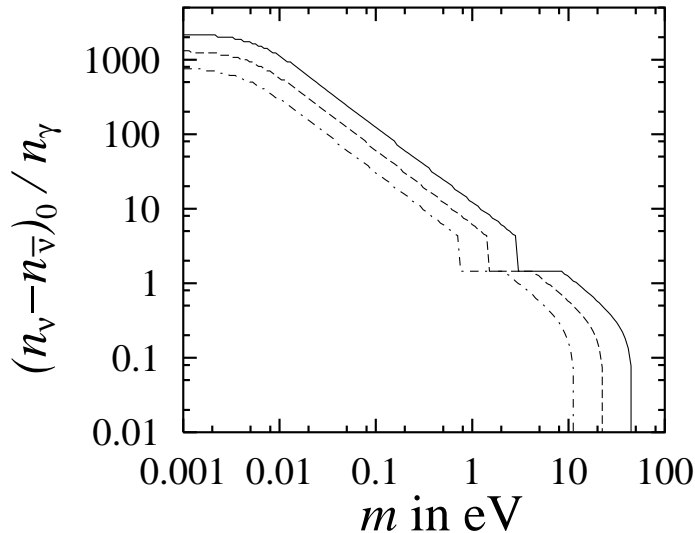}}
\caption[]{Neutrino mass and degeneracy allowed by cosmological energy
density bounds. The curves are contours for $h^2\Omega F_\nu=\frac18,\
\frac14$ and $\frac12$ respectively. From Ref.~\cite{KP99}.}
\label{f:KP}
\end{center}
\end{figure}
%%%%%%%%%%%%%%%%%%%%%%%%%
Before discarding light neutrinos as dark matter candidates on this
ground, I would like to emphasize that there is a very strong
assumption implicit in this entire argument, viz.\ that the neutrinos
have no chemical potential so that their number is the same as that of
the antineutrinos. If chemical potential is nonzero, the number
density of neutrinos depend on it as well as the temperature, and one
cannot even start with the simple relation of Eq.\ (\ref{3/4}).

If the neutrinos have chemical potential, they can contribute a larger
fraction to the energy density of the universe. This issue has been
analyzed recently \cite{KP99}. We show the results in
Fig.~\ref{f:KP}. The horizontal axis corresponds to neutrino mass. The
vertical axis is $n_\nu-n_{\bar\nu}$ in units of the CMBR $n_\gamma$
in the present universe. The lines in this plot correspond to the
values $h^2\Omega F_\nu=\frac18,\ \frac14$ and $\frac12$ respectively,
starting from the inner line.

%%%%%%%%%%%%%%%%%%%%%%%%%
\section{Axions}
%%%%%%%%%%%%%%%%%%%%%%%%%
So far we talked about light neutrinos. They are known to exist. The
other candidates to be discussed have not yet been detected in any
experiment. Thus, I need some motivation for talking about them.

For the case at hand, the motivation comes from strong
interactions. These interactions are believed to be described by the 
gauge theory called QCD. In a gauge theory, one decides on some
symmetry of the Lagrangian and writes down all terms that are
cosistent with this symmetry. For strong interactions, this symmetry
was found in the 1970s. If one cares to do only perturbative
calculations with the Lagrangian obtained from this symmetry, there is
no problem. If, however, one includes non-perturbative effects, one
encounters one term which violates parity and time-reversal
symmetries:
\begin{eqnarray}
{\cal L}_{\rm QCD} = {\cal L}_{\rm pert} + {\theta g^2 \over 32\pi^2}
G_{\mu\nu} \widetilde G_{\mu\nu} \,,
\label{thGG}
\end{eqnarray}
where $G_{\mu\nu}$ is the generalization of the electromagnetic
field-strength tensor for the relevant symmetry. Here $g$ is the gauge
coupling constant, and $\theta$ is an arbitrary parameter.

The term involving this parameter is the generalization of a term
$\vec E \cdot \vec B$, which violates parity and time-reversal
symmetries. Among other things, this term will give rise to an
electric dipole moment for the neutron. The experimental bounds are
very strong on this, which restricts this parameter severely:
$\theta<10^{-10}$.

Very small parameters always present certain conceptual difficulties
in quantum field theories. If they are not zero, one has to justify
them by some symmetry. Peccei and Quinn discovered such a symmetry by
introducing a Higgs field $\widehat a(x)$ which has an interaction
\begin{eqnarray}
-\; {g^2 \over 32\pi^2} \; {\widehat a(x) \over f_{\rm PQ}} \;
G_{\mu\nu} \widetilde G_{\mu\nu} \,,
\label{aGG}
\end{eqnarray}
where $f_{\rm PQ}$ has the dimensions of mass. The field, of course,
has other interactions, including a potential of its own, which 
is minimum at $\widehat a=\theta f_{\rm PQ}$. In general, now, we can
write $\widehat a(x)=\theta f_{\rm PQ}+a(x)$. This field $a(x)$
vanishes at the minimum and is qualified to be a quantum field. It is
called the {\em axion field}.

Notice how it has solved our original problem. If we add the terms
shown in Eqs.\ (\ref{thGG}) and (\ref{aGG}) and use the definition of
the quantum field, we see that the parameter $\theta$ cancels
out. Said another way, $\theta$ is forced to equal to zero by this
procedure. We have an interaction of the axion obtained by replacing
$\widehat a$ by $a$ in Eq.\ (\ref{aGG}), but that does not violate
parity and time reversal provided the axion is intrinsically negative
under these symmetries.

We now examine how the axion can be important as the dark matter. The
equation of motion of the axion field, neglecting its interactions, is
\begin{eqnarray}
\ddot a + 3H \dot a + m_a^2 a = 0 \,,
\label{aeq}
\end{eqnarray}
assuming the field to be spatially homogeneous. This is similar to the
Klein-Gordon equation, but there are two differences. One of them is
the second term on the left side, which shows the effects of the
expansion of the universe. The other is hidden in the notation, and is
the fact that the `mass' $m_a$ appearing in the equation is not a
constant. It arises due to co-operative effects, and depends on the
temperature. 

At early times, when the temperature was very high, $m_a$ was
zero. This is because all non-perturbative effects involve a factor
$\exp(-1/g^2)$, which is negligible since at high energies $g$ is very
small for QCD. In this stage, the solution of Eq.\ (\ref{aeq}) is
given by $a={\rm constant}$. As the universe cools down, $g$
increases, and becomes large enough to make the non-perturbative
effects important when the temperature drops down sufficiently, say
below a value $T_{\rm QCD}$. To find the solution after this time, let
us first forget about $H$ and also the time-dependence of $m_a$. In
that case, the solution would be
\begin{eqnarray}
a = A \cos m_at \,,
\end{eqnarray}
where $A$ is a constant. If $H\ll m_a$ and also the time variation of
$m_a$ is small, we can still try a solution like this, where now both
$A$ and $m_a$ should be regarded as functions of time. Substituting
the solution in Eq.\ (\ref{aeq}), we now obtain
\begin{eqnarray}
{d\over dt} (m_a A^2) = -3H m_aA^2 \,,
\end{eqnarray}
or
\begin{eqnarray}
m_a A^2 R^3 = {\rm constant},
\label{constant}
\end{eqnarray}
where $R$ is the scale factor of the universe, i.e., $H=\dot R/R$. The
energy density in these oscillations, at the present time, would be
given by
\begin{eqnarray}
\rho_0 =  \frac12 \left( m_a^2 A^2 \right)_0 \,.
\end{eqnarray}
The axion mass in the present universe is determined by $f_{\rm
PQ}$. It depends mildly on the axion models, and is given by
\begin{eqnarray}
(m_a)_0 f_{\rm PQ} \simeq m_\pi f_\pi \,,
\label{curralg}
\end{eqnarray}
where the quantities on the right sides relate to properties of the
pions, and are very well-known. However, we don't know the magnitude
of the oscillations in the present universe. Therefore, we use Eq.\
(\ref{constant}) to write
\begin{eqnarray}
\rho_0 =  \frac12 \left( m_a \right)_0 \left( m_aA^2 \right)_T \times
(R^3/R_0^3) \,, 
\end{eqnarray}
where $R$ is the value of the scale factor at any arbitrary
temperature $T$. The oscillations started, as we said, around $T\simeq
T_{\rm QCD}\simeq 100$\,MeV. At that time, $m_a\simeq H\simeq
T^2/M_P$, and $A\simeq f_{\rm PQ}$. Using these estimates and
utilizing the fact that $R/R_0=T_0/T$, we obtain
\begin{eqnarray}
\rho_0 \simeq   \frac12 {m_\pi f_\pi \over f_{\rm PQ}} \cdot \left(
{T_{\rm QCD}^2 f_{\rm PQ}^2 \over M_P} \right) \cdot \left(
{T_0 \over T_{\rm QCD}} \right)^3 \,,
\end{eqnarray}
which gives
\begin{eqnarray}
\Omega_a \simeq 10^{-12} {f_{\rm PQ} \over 1\,{\rm GeV}} \,.
\end{eqnarray}
Thus, axions can be an important source of dark matter if $f_{\rm PQ}$
is close to $10^{12}$\,GeV.

%%%%%%%%%%%%%%%%%%%%%%%%%
\begin{figure}
\begin{center}
\centerline{\epsfysize=8cm %\epsfxsize=\textwidth
\epsfbox{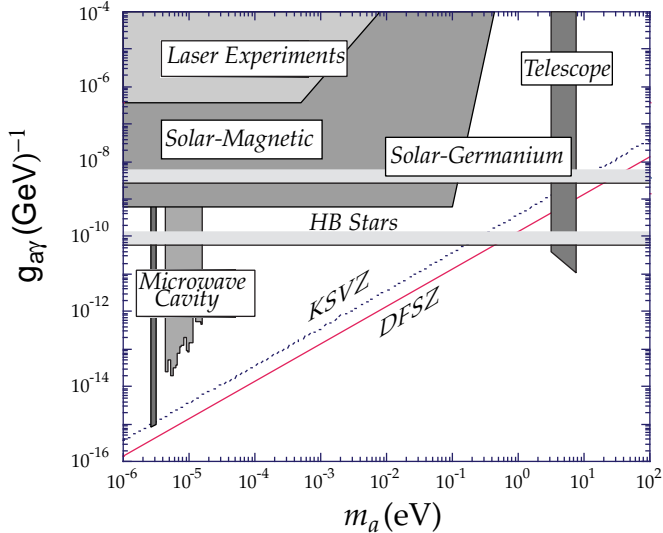}} 
\caption[]{Summary of axion search results. The shaded regions are
excluded. Taken from Ref.~\cite{PDG}. Some of the legends have been
slightly edited to conform with the notations used in this paper.}
\label{f:axsearch}
\end{center}
\end{figure}
%%%%%%%%%%%%%%%%%%%%%%%%%
Let us now see what is experimentally known about these axions. In
Fig.~\ref{f:axsearch}, we show the present status \cite{PDG}. The
horizontal axis is the axion mass. The vertical axis denotes its
coupling with two photons, which allows the decay mode
$a\to\gamma\gamma$. As I said earlier, there is slight model
dependence in these quantities. The predictions of two compelling
models are shown in the graph by the lines marked DFSZ and KSVZ.

If the axion mass is in the eV range, the decay photons could be
detected through telescopes, and the failure to do so has been
indicated in the figure. However, if $f_{\rm PQ}$ indeed equals
$10^{12}$\,GeV and we take Eq.\ (\ref{curralg}) as a strict equality,
the axion mass comes out to be $10^{-5}$\,eV. So, if the axions has to
contribute significantly to the energy density of the universe, we
should find them around that value of mass. In this case, the two
photon decay mode is so slow that the axion can be considered stable
on cosmological time scales. However, because of the coupling, the
axions may be resonantly converted into a microwave signal in a cavity
where a strong magnetic field is present. These experiments are very
promising. As we see, they are already probing very close to the
interesting region.

%%%%%%%%%%%%%%%%%%%%%%%%%
\section{WIMPs}
%%%%%%%%%%%%%%%%%%%%%%%%%
We now come to the WIMPs, which is an acronym for ``Weakly Interaction
Massive Particles.'' There are a variety of ways in which such
particles may be motivated. The most recent and most compelling of
them is supersymmetry. Supersymmetric models have new particles ---
bosonic partners of known fermions and fermionic partners of known
bosons. There are phenomenolgically strong reasons to suspect that
these new particles, or ``superpartners'', can be produced or
annihilated only in pairs. If that is true, the lightest of these
superpartners will be a stable particle. It will not decay
spontaneously. These can be examples of WIMPs. There are other models
also which predict WIMPs.

The number density of these WIMPs in the early universe is governed by
the Boltzmann equation. In the homogeneous and isotropic universe,
this takes the form
\begin{eqnarray}
{dn\over dt} = -3Hn - \sigma v (n^2 - n_{\rm eq}^2) \,,
\end{eqnarray}
where $n_{\rm eq}$ is the equilibrium number density for the
temperature at any given time. The decrease of number density due to
the expansion of the universe is represented in the first term in this
equation, and the same due to pair annihilations goes in the second
term. 

%%%%%%%%%%%%%%%%%%%%%%%%%
\begin{figure}
\begin{center}
\centerline{\epsfysize=8cm %\epsfxsize=\textwidth
\epsfbox{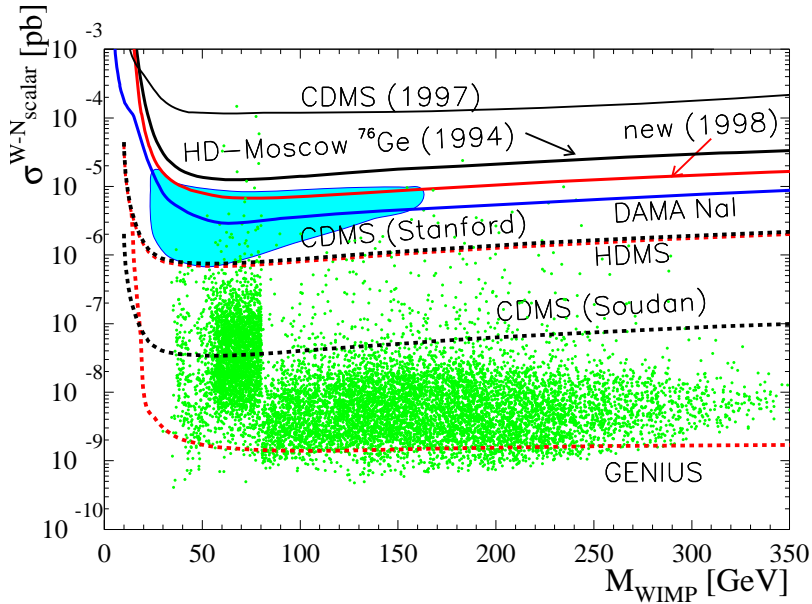}} 
\caption[]{Summary of WIMP search results and future prospects. From
Ref.~\cite{Baudis99}} 
\label{f:klapdor}
\end{center}
\end{figure}
%%%%%%%%%%%%%%%%%%%%%%%%%
It is convenient to switch to dimensionless variables defined by
\begin{eqnarray}
y = {n \over s} \,, \qquad x = {m/T}\,,
\end{eqnarray}
where $m$ is the mass of the WIMP and $s$ is the entropy of the
background photons at a temperature $T$. The evolution equation now
becomes
\begin{eqnarray}
{dy\over dx} = - {y_{\rm eq} \over x} {\Gamma_{\rm ann} \over H(x)}
\left[ \left( {y \over y_{\rm eq}} \right)^2 -1 
\right] \,,
\label{yeqn}
\end{eqnarray}
where $\Gamma_{\rm ann}=n_{\rm eq}\sigma v \sim m^3
x^{-3/2}e^{-x}\sigma v$, putting in the equilibrium number density for
a non-relativistic particle. The important thing to notice here is
that $y$ hardly changes if $\Gamma_{\rm ann}\ll H(x)$ at any given era
characterized by a temperature determined by $x$. Using $H(x)\sim
T^2/M_P = m^2/M_Px^2$, we find
\begin{eqnarray}
{\Gamma_{\rm ann} \over H(x)} \sim m M_P x^{1/2}e^{-x}\sigma v \,.
\label{G/H}
\end{eqnarray}
As $x\to \infty$, i.e., for large time, this ratio goes to zero. It
means that after a while, the pair annihilation rate will be
negligible, because the universe has meanwhile become large enough so
that the particles do not find each other. The remnant density of the
WIMPs can be obtained by numerical integration of Eq.\ (\ref{yeqn}),
but a rough estimate can be obtained by the density at the era when
$\Gamma_{\rm ann} = H(x)$. The solution of $x$ Eq.\ (\ref{G/H}) is now
determined once the mass and the annihilation cross section are
known. After this era, the number density just scales inversely as the
volume of the universe.

Let me now turn to the experimental searches \cite{PhRep}. These are
summarized it in Fig.~\ref{f:klapdor}. As I just said, the parameters
which determine the density of WIMPs in the present universe are the
mass and the annihilation cross section. The two axes in this plot are
precisely these two parameters. For the cross section, only the cross
section with the nucleons is probed in the experiments. The solid
lines are exclusions plots from already published data, and the dotted
lines are projections for future experiments. The scatter plots are
expected values of the parameters for various choices of the basic
parameters in the supersymmetric model. It does seem that the
experiments in the near future will be able to decide whether WIMPs
really dominate the energy density of the universe.

%%%%%%%%%%%%%%%%%%%%%%%%%

%%%%%%%%%%%%%%%%%%%%%%%%%

\end{document}